\documentclass{egpubl}
\usepackage{eg2022}
 
\ConferencePaper        
\usepackage[T1]{fontenc}
\usepackage{dfadobe}  

\usepackage{cite}  
\BibtexOrBiblatex
\electronicVersion
\PrintedOrElectronic

\ifpdf \usepackage[pdftex]{graphicx} \pdfcompresslevel=9
\else \usepackage[dvips]{graphicx} \fi

\usepackage{egweblnk} 

\newcommand*\samethanks[1][\value{footnote}]{\footnotemark[#1]}

\title[CLIP-based Neural Neighbor Style Transfer for 3D Assets]%
      {CLIP-based Neural Neighbor Style Transfer for 3D Assets}

\author[Shailesh Mishra \& Jonathan Granskog]
{\parbox{\textwidth}{\centering Shailesh\,Mishra\thanks{Work done while employed at NVIDIA}$^{1,2}$
        and Jonathan Granskog\samethanks$^{2}$
        }
        \\
{\parbox{\textwidth}{\centering $^1$Saarland University, Germany\\
         $^2$NVIDIA
       }
}
}

\begin{document}

\teaser{
 \includegraphics[width=1.0\linewidth]{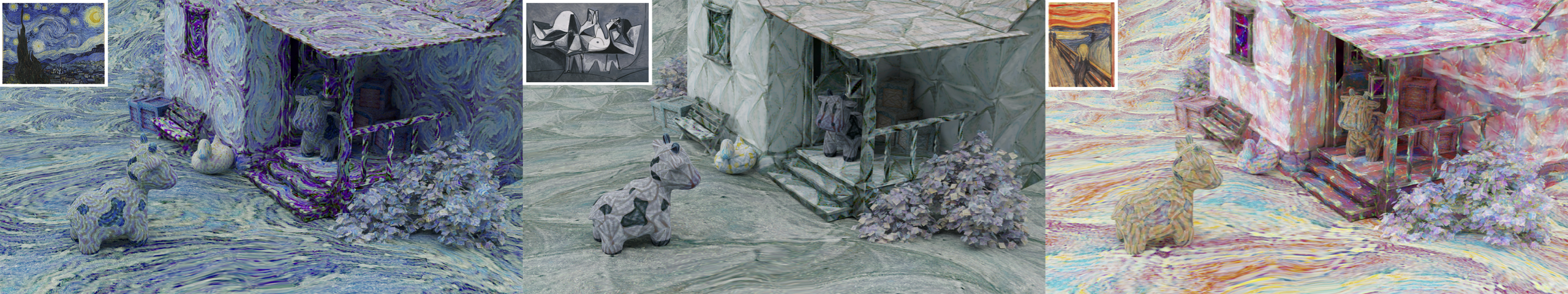}
 \centering
  \caption{Our method can stylize objects such that they can be rendered easily with traditional 3D renderers. We show renders of a scene containing objects independently processed by our method using different style images (shown in the top left corner of each render).}
\label{fig:teaser}
}

\maketitle
\begin{abstract}
   We present a method for transferring the style from a set of images to a 3D object. The texture appearance of an asset is optimized with a differentiable renderer in a pipeline based on losses using pretrained deep neural networks. More specifically, we utilize a nearest-neighbor feature matching loss with CLIP-ResNet50 to extract the style from images. We show that a CLIP-based style loss provides a different appearance over a VGG-based loss by focusing more on texture over geometric shapes. Additionally, we extend the loss to support multiple images and enable loss-based control over the color palette combined with automatic color palette extraction from style images. 

\ccsdesc[300]{Computing methodologies~Appearance and texture representations}
\ccsdesc[300]{Computing methodologies~Rasterization}
\ccsdesc[300]{Computing methodologies~Supervised learning by regression}
\ccsdesc[100]{Applied computing~Media arts}

\end{abstract} 

\section{Introduction}

As physically-based rendering becomes ubiquitous, artistic style reveals itself as a key factor of differentiation from other works. Modern animated shows often use manual stylization to produce memorable visual qualities (see Spiderverse \cite{spiderverse} or Arcane \cite{arcane}). As style becomes an increasingly important factor in drawing the attention of audiences, manual processes might become too tedious especially for large asset libraries. Automatic methods can also provide great starting points for stylization.

Our research presents an initial attempt to tackle automatic stylization of textures on 3D assets based on a specific style described by a set of images (\autoref{fig:teaser}). We motivate the choice of image-based stylization, rather than textual descriptors, because productions often start with an initial collection of 2D concept art that reflect the desired style. Our approach aims to extract the style from these images using differentiable rendering and pretrained deep neural networks.

\section{Related work}
\vspace{-1mm}
In the image domain, neural style transfer learns the style of an input image by minimizing the distance between statistics of feature maps, such as Gram matrices, in a pretrained convolutional neural network \cite{Gatys:2016}. Many recent works have improved on this framework; see \cite{Jing:2020} for a comprehensive review of neural style transfer. 

\begin{figure*}[ht]
    \small
    \begin{tabular}{ccccccc}
    Style image & Original & CLIP Embedding & VGG Gram & CLIP Gram  & VGG NNFM & CLIP NNFM (Ours) \\
    \includegraphics[clip,trim=0 0 0 0,width=0.12\linewidth]{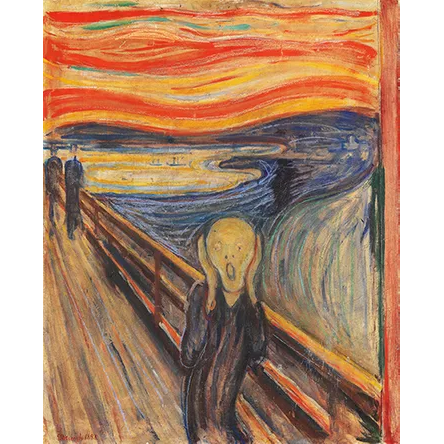} &
    \includegraphics[clip,trim=100 100 100 100,width=0.12\linewidth]{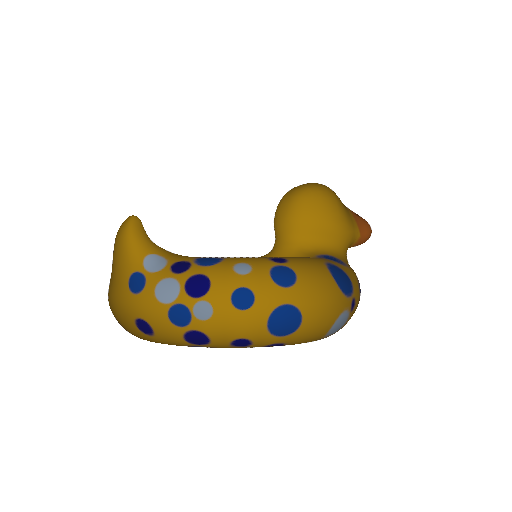} &
    \includegraphics[clip,trim=100 100 100 100,width=0.12\linewidth]{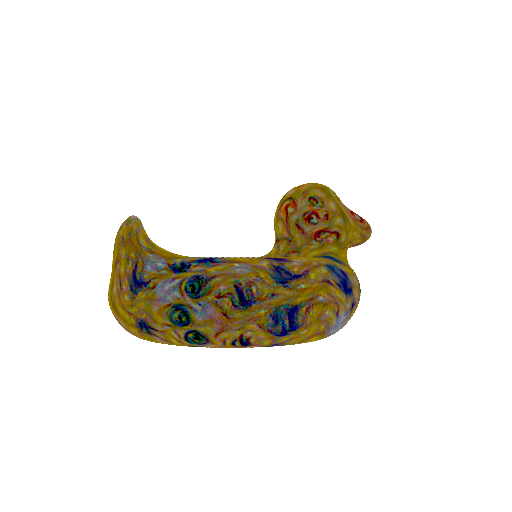} &
    \includegraphics[clip,trim=100 100 100 100,width=0.12\linewidth]{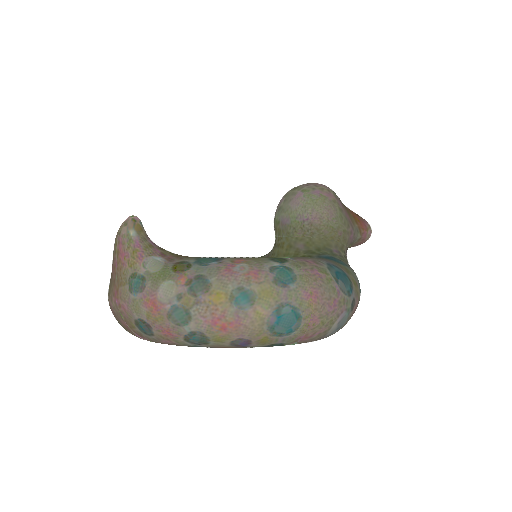} &
    \includegraphics[clip,trim=100 100 100 100,width=0.12\linewidth]{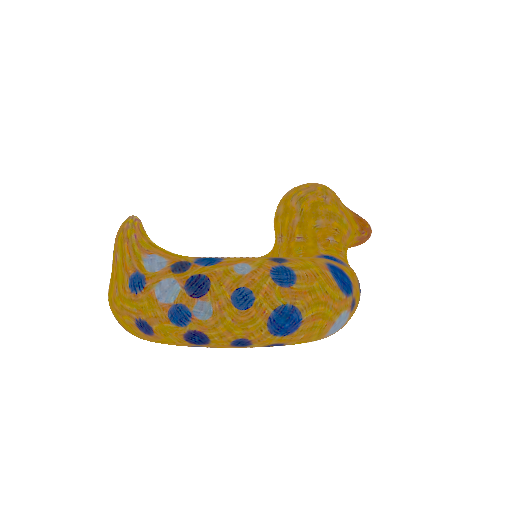} &
    \includegraphics[clip,trim=100 100 100 100,width=0.12\linewidth]{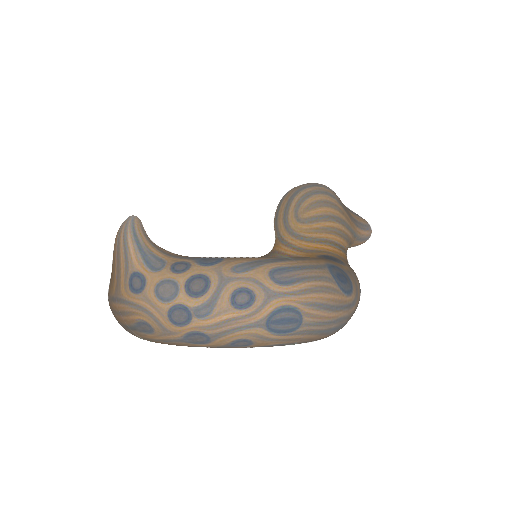} &
    \includegraphics[clip,trim=100 100 100 100,width=0.12\linewidth]{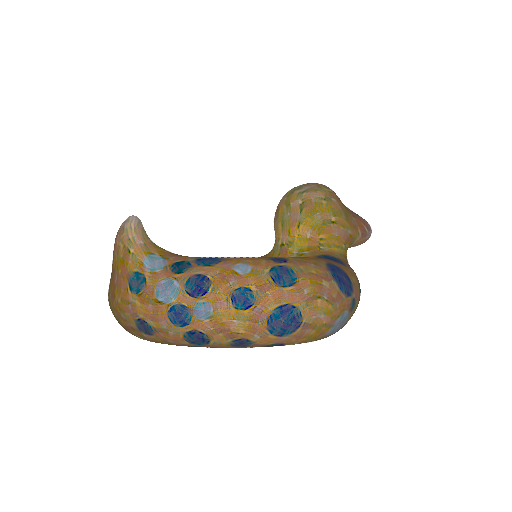} \\
    \includegraphics[clip,trim=0 0 0 0,width=0.12\linewidth]{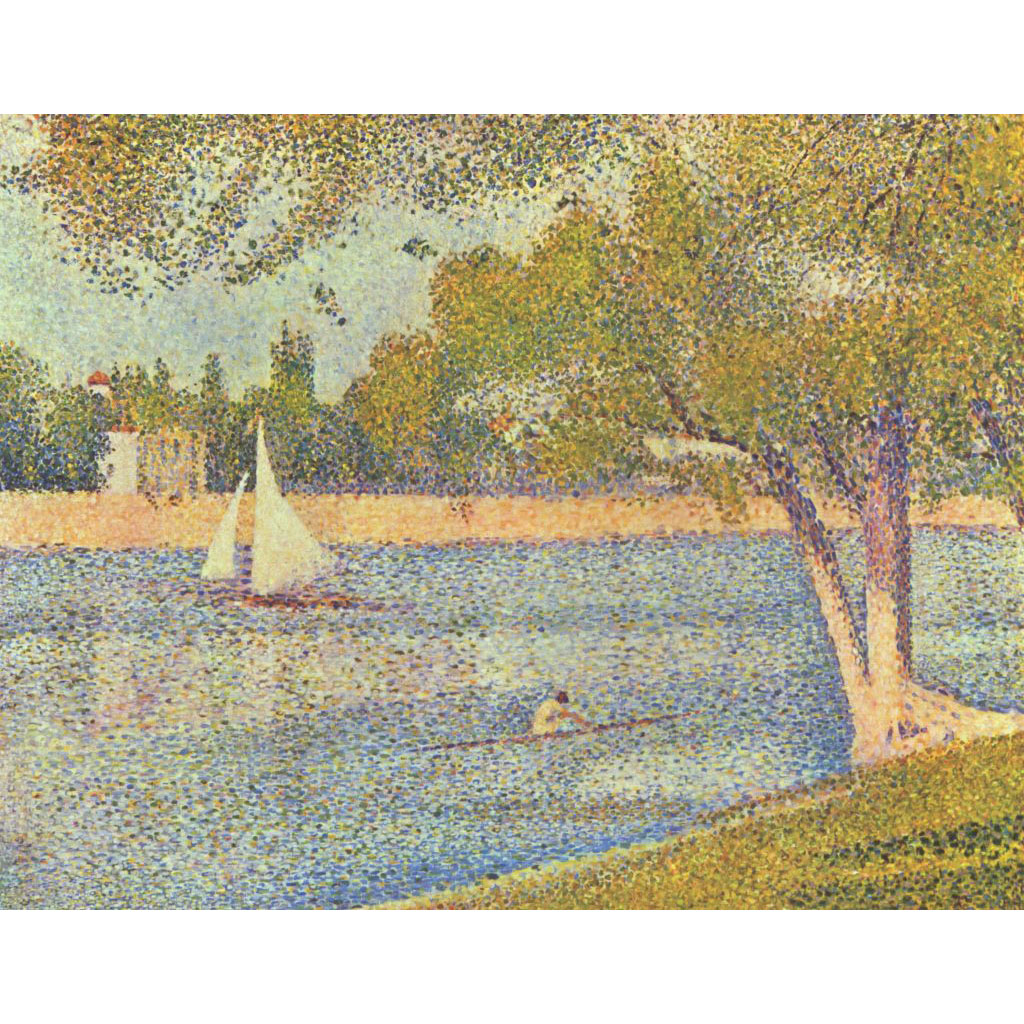} &
    \includegraphics[clip,trim=100 100 100 100,width=0.12\linewidth]{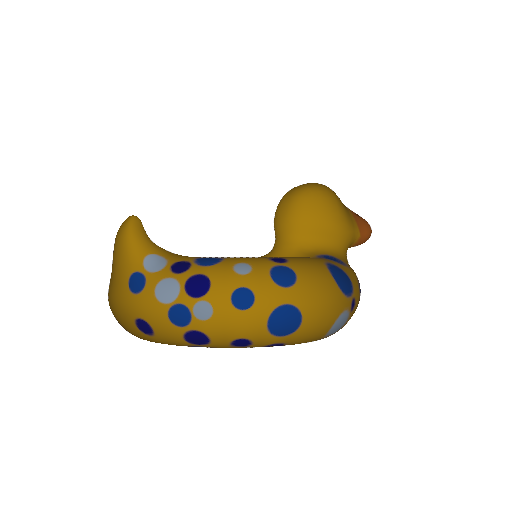} &
    \includegraphics[clip,trim=100 100 100 100,width=0.12\linewidth]{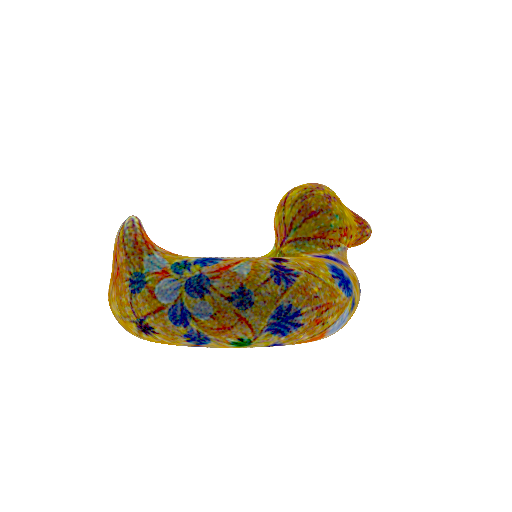} &
    \includegraphics[clip,trim=100 100 100 100,width=0.12\linewidth]{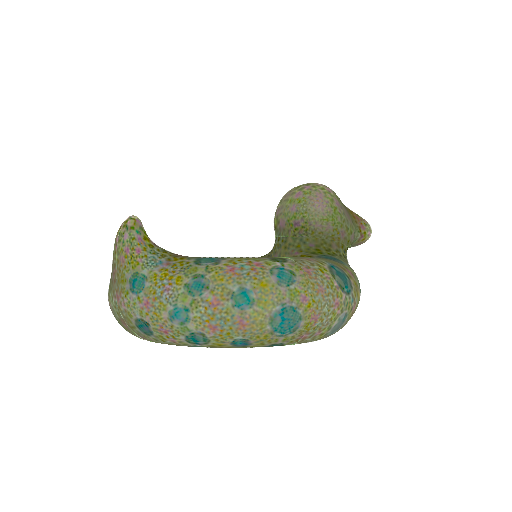} &
    \includegraphics[clip,trim=100 100 100 100,width=0.12\linewidth]{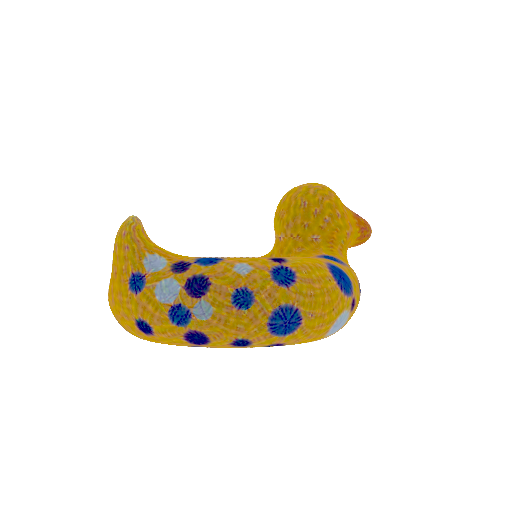} &
    \includegraphics[clip,trim=100 100 100 100,width=0.12\linewidth]{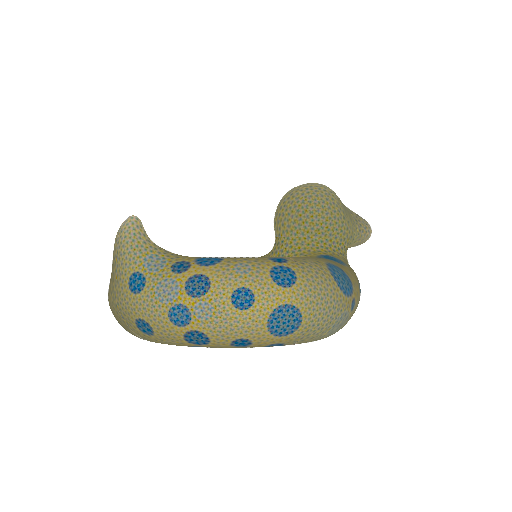} &
    \includegraphics[clip,trim=100 100 100 100,width=0.12\linewidth]{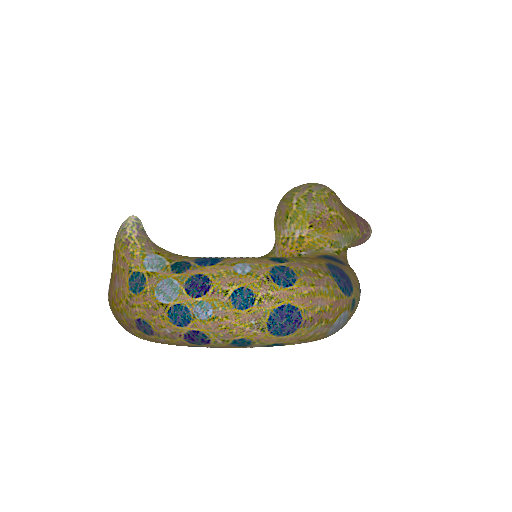} \\
    \end{tabular}
    \centering
    \caption{
        \label{fig:transfer-comparison}
        Comparison of different style losses without our color palette loss. Directly using the CLIP embedding also extracts content from the style image, in this case eyes and leaves, and prevents functioning transfer of style. Both Gram matrix-based methods show poor transfer of style whereas the NNFM-based methods work better. However, VGG and CLIP extract different appearances from the style image. CLIP focuses more on texture whereas VGG captures lines and points. 
    }
\end{figure*}

We optimize the styles of textures on 3D assets with differentiable rendering \cite{Laine:2020} and features from a pretrained CLIP-ResNet50 network \cite{Radford:2021}, improving textural detail compared to VGG16 \cite{Liu:2015}. Instead of Gram matrices, which do not work well with CLIP, we follow the nearest-neighbor feature matching approach presented by Kolkin et al. \cite{Kolkin:2022}, which was later extended to neural radiance fields \cite{Zhang:2022}. CLIP, which shares a latent space for image and text input, has proven powerful in both stylizing and generating images, meshes and neural fields using text prompts \cite{Jain:2020,Kwon:2021,Michel:2021,Ramesh:2022}. Due to its proven ability to represent style, we evaluate and show that CLIP-ResNet50 can bring benefits to neural style transfer for textures when using a nearest-neighbor feature loss (see \autoref{fig:transfer-comparison}). 

\begin{figure*}[ht]
    \includegraphics[width=1.0\linewidth]{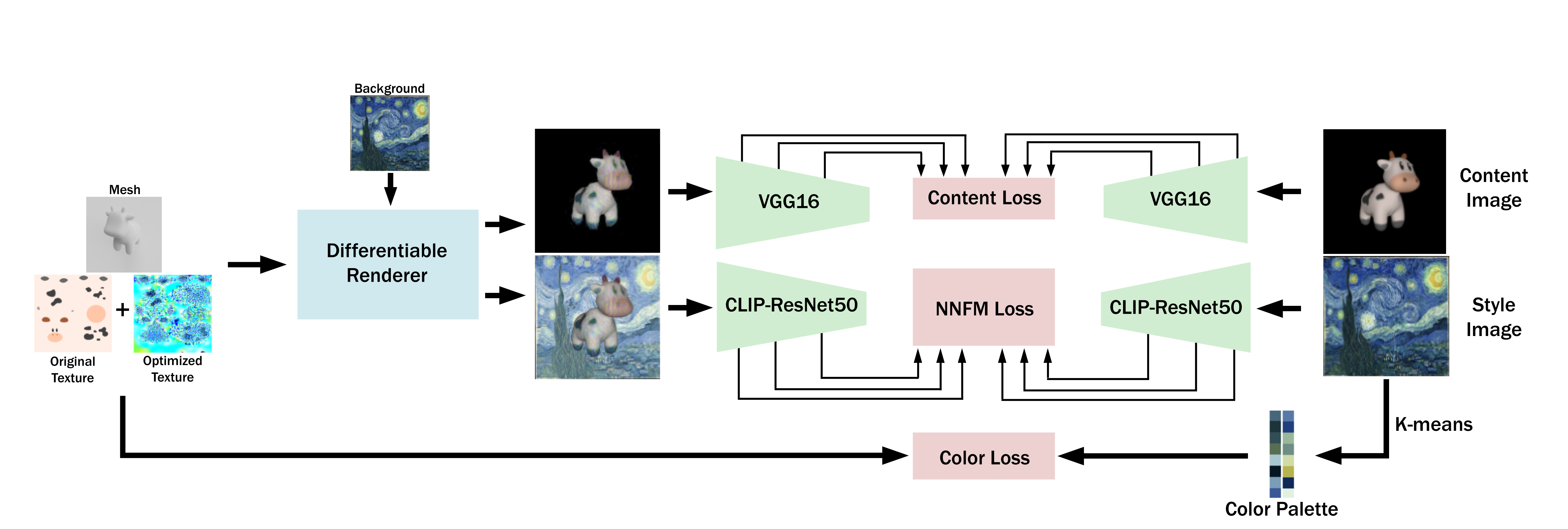}
    \centering
    \caption{
        \label{fig:pipeline}
        Our pipeline begins by rendering an object twice; once with background and once without. The render without background produces the VGG16 content loss whereas the other image is compared to the style image with the nearest-neighbor feature matching style loss. An additional color loss applied directly to the texture constrains the optimization towards an extracted color palette. 
    }
\end{figure*}


\section{Method}
\label{sec:method}

In this section, we describe our method for transferring the style from a set of images to the texture $\mathcal{T}$ of a 3D asset. To match the style of an input image $\mathcal{S}$, we use a nearest-neighbor feature matching (NNFM) loss that minimizes distances in feature space while ignoring spatial locations \cite{Kolkin:2022}. For texture optimization, we rely on the differentiable renderer from Laine et al. \cite{Laine:2020} due to its simplicity, but buffers output by a deferred renderer could also be used. See \autoref{fig:pipeline} for an overview of our approach and the supplementary material for more details. 

We present three main contributions; an extension of the nearest-neighbor feature matching (NNFM) style loss to CLIP-ResNet50 and multiple style images, and a color palette loss to allow additional artistic control and improve results. Our final loss consists of three components
\vspace{-1mm}
\begin{equation}
    \label{eq:losses}
    L = \lambda_{nnfm} L_{nnfm} + \lambda_{content} L_{content} + \lambda_{color} L_{color}
\end{equation}

For additional artistic control, we optimize a separate style texture $\mathcal{T_S}$, initialized to zero, and keep the original texture $\mathcal{T}$ fixed. The two are then added and clamped before rendering. We use batched training, and for each batch element, we randomize the camera and lighting, and render the object with the texture. For more optimization details, see the supplementary material.

\paragraph*{NNFM style loss} 

Here, we describe our application of the NNFM style loss \cite{Kolkin:2022}. Given a render output image $\mathcal{I}$ and style image $\mathcal{S}$, we forward propagate through a pretrained neural network and extract feature maps from a layer $L$, and reshape the feature maps into sets of $N$ and $M$ feature vectors $F^\mathcal{S}$ and $F^\mathcal{I}$ respectively. Then, we minimize the distance of each render feature vector to the nearest style feature vector according to

\vspace{-1.5mm}
\begin{equation}
L_{nnfm} = \frac{1}{M} \sum^{M}_{i} \min^{N}_{j} \mathcal{D}(F^{\mathcal{I}}_i, F^{\mathcal{S}}_j)
\end{equation}

where $\mathcal{D}$ is the cosine distance between the feature vectors. To account for multiple style images $\lbrace\mathcal{S}_k\rbrace_{k=1...K}$, we expand the set of style feature vectors by concatenating the feature vectors extracted from each style image $F^{\mathcal{S}} = F^{\mathcal{S}_1} \oplus F^{\mathcal{S}_2} ... \oplus F^{\mathcal{S}_K}$ where $\oplus$ denotes concatenation.

For CLIP-ResNet50, we use the outputs of the second convolutions from each block in layer3 and layer4 to compute the loss. We also follow Wang et al. \cite{Wang:2021} and smooth the features with a Softmax before computing the style loss and, to aid style transfer, we set the style image as the background image during feature extraction. 

\paragraph*{Content loss} To retain the characteristic content of the original texture, we utilize a standard VGG16-based content loss which minimizes the L2 distance between feature maps of a content image $\mathcal{C}$ and input image $\mathcal{I}$: $L_{content} = L_2(F^{\mathcal{C}}, F^{\mathcal{I}})$. We compute the loss on features from layers 11, 13 and 15 in VGG16, and we do not include the style image as background as in \autoref{fig:pipeline}. 

\paragraph*{Color palette loss}

CLIP struggles to match colors in the style image accurately without an additional color loss term to direct the texture stylization (\autoref{fig:loss-ablation}). We include a loss that matches texture colors to a color palette $P$ of size $Q$, either automatically extracted from the style image or user-provided for artistic control. For each pixel in the summed texture $\mathcal{T_S} + \mathcal{T}$, we minimize the L2 distance to the nearest RGB color in the palette
\vspace{-0.5mm}
\begin{equation}
L_{color} =  \frac{1}{WH} \sum^{W, H}_{i, j} \min^Q_q L_2((\mathcal{T}_S + \mathcal{T})_{i,j}, P_q)
\end{equation}

We apply the loss to the texture, rather than the output render $\mathcal{I}$, to ignore the effect of shading and background. To automatically extract the color palette from an image, we cluster colors based on K-means and the final cluster centers become the palette. With multiple style images, we select a primary style image for the color palette extraction, but other methods would also work well. 

\section{Results}

In this section, we present results from our image-based style transfer method for textures of 3D assets. We compare each style transfer approach and the impact of each loss term. Then, we show artistic control of colors and the benefit of multiple style images. Additional comparisons can be found in the supplemental material. 

\paragraph*{Style transfer comparison}

\begin{figure*}[ht]
    \small
    \begin{tabular}{ccccc}
    Style & K-means (VGG16) & Manual (VGG16) & K-means (CLIP-ResNet50) & Manual (CLIP-ResNet50) \\
    \includegraphics[clip,trim=0 0 0 0,width=0.18\linewidth]{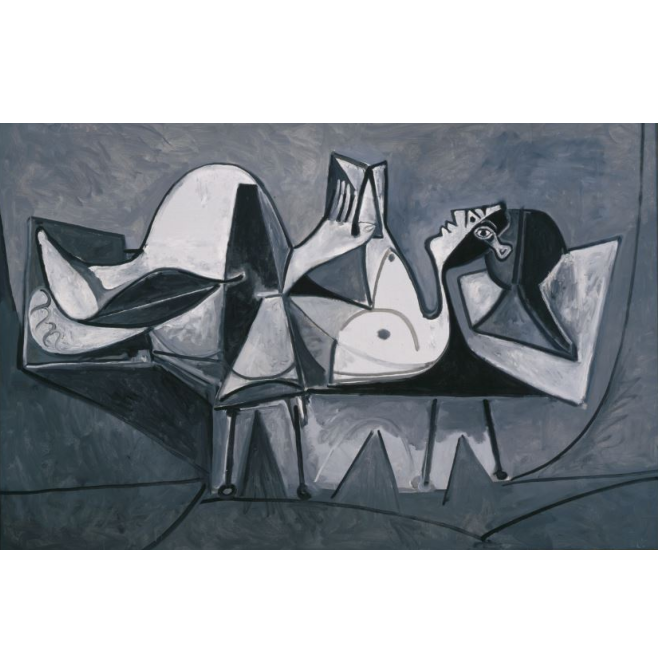} &
    \includegraphics[clip,trim=50 50 50 50,width=0.18\linewidth]{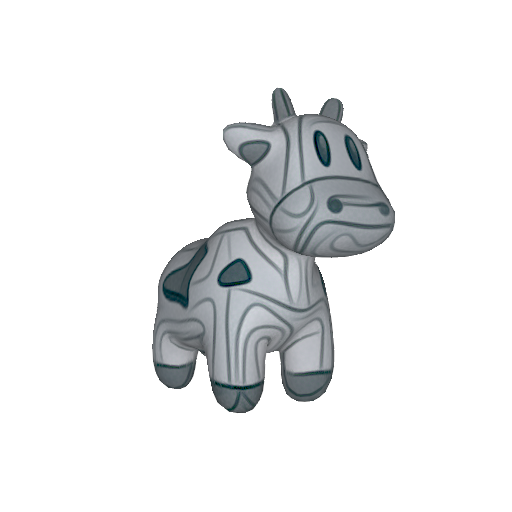} &
    \includegraphics[clip,trim=50 50 50 50,width=0.18\linewidth]{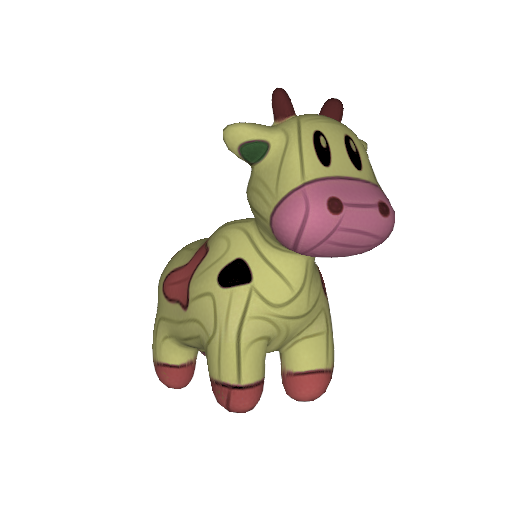} &
    \includegraphics[clip,trim=50 50 50 50,width=0.18\linewidth]{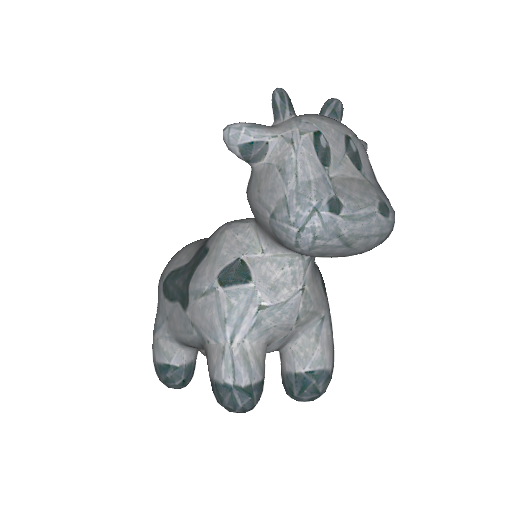} &
    \includegraphics[clip,trim=50 50 50 50,width=0.18\linewidth]{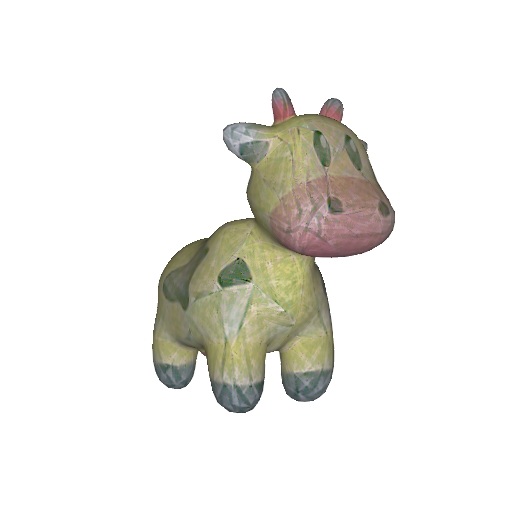} \\
    \end{tabular}
    \centering
    \caption{
        \label{fig:color-loss}
        Our color palette loss allows controlling the color of the end result, even if the selected palette is very different from the colors found in the style image. Columns labeled manual use a manually-selected color palette and the other columns use K-means extraction. One can also see that CLIP-ResNet50 focuses more on texture whereas VGG16 prefers synthesizing geometric features such as lines. 
    }
\end{figure*}

In \autoref{fig:transfer-comparison}, we show a comparison of various style transfer methods. A direct loss on the CLIP embedding matches both content and style and as such eyes and leaves are synthesized in the textures. The standard Gram-matrix-based approach with both VGG16 \cite{Gatys:2016} and CLIP-ResNet50 \cite{Radford:2021} is unable to properly learn a global style appearance whereas the nearest-neighbor-based VGG16 approach \cite{Kolkin:2022} manages to capture brush strokes and geometric shapes well. CLIP-ResNet50 NNFM  focuses instead more on learning a textural appearance. For example, in the top row, VGG NNFM learns long brush strokes flowing along the mesh but lacks texture whereas CLIP learns a painterly appearance but is missing long strokes. See also \autoref{fig:color-loss}. For VGG16, we follow Zhang et al. \cite{Zhang:2022} and use feature maps from layers 11, 13 and 15.

\begin{figure}[ht]
    \small
    \begin{tabular}{cccc}
    Style & All losses & No color loss & NNFM only \\
    \includegraphics[clip,trim=450 0 450 0,width=0.21\linewidth]{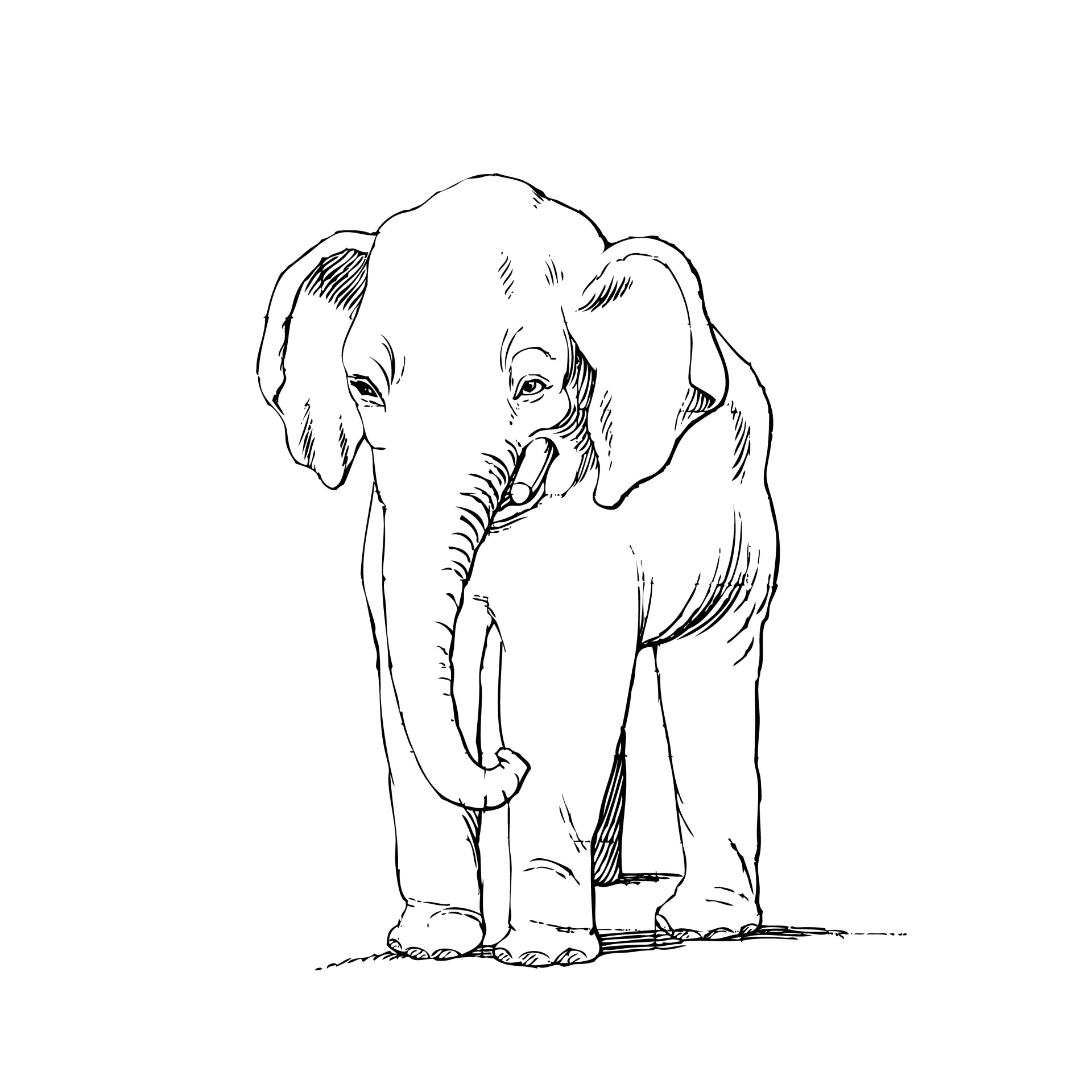} &
    \includegraphics[clip,trim=125 0 100 0,width=0.21\linewidth]{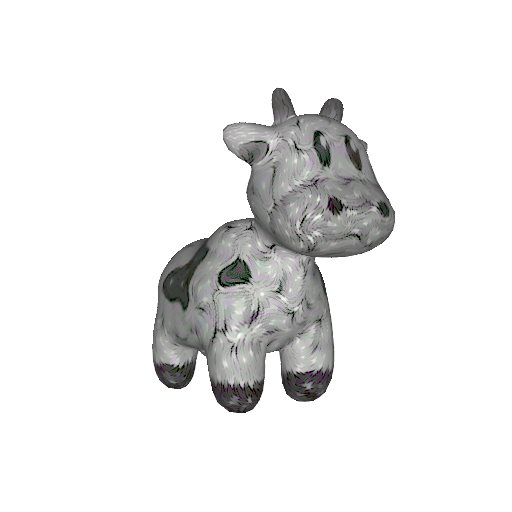} &
    \includegraphics[clip,trim=125 0 100 0,width=0.21\linewidth]{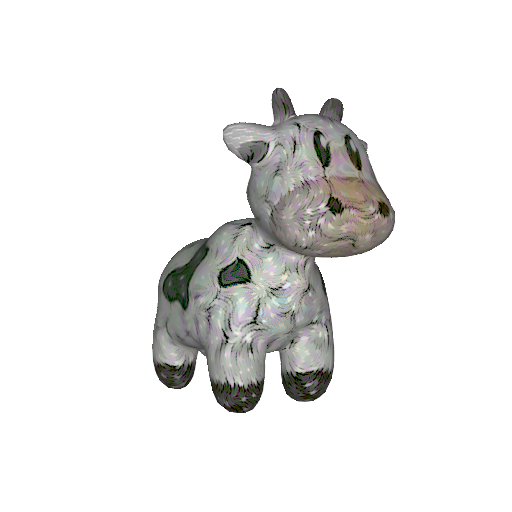} &
    \includegraphics[clip,trim=125 0 100 0,width=0.21\linewidth]{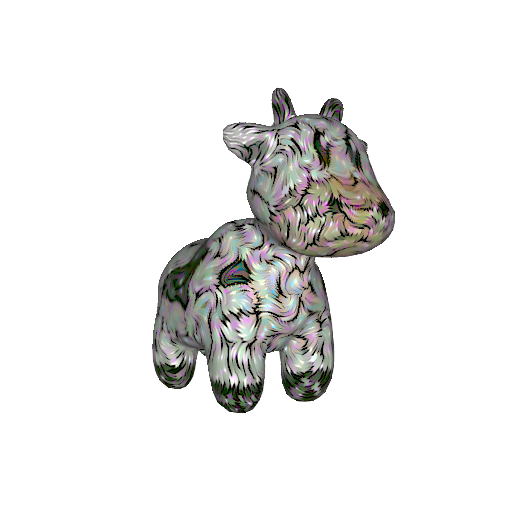} \\
    \end{tabular}
    \centering
    \caption{
        \label{fig:loss-ablation}
        The impact of each loss in CLIP-based NNFM style transfer when using a black and white line art drawing as the style. The color loss ensures colors match the style image whereas the content loss ensures the content is preserved.
    }
\end{figure}

\paragraph*{Loss ablation}

Our style transfer method consists of three losses; a NNFM style loss, a VGG16-based content loss and a color palette loss (\autoref{sec:method}). \autoref{fig:loss-ablation} shows the impact of each loss on our style transfer method. The content loss constrains the optimization such that we do not lose characteristic features from the original texture, such as eyes of the cow. The automatic palette extraction, based on K-means clustering, combined with the color palette loss help match the colors of the style image and \autoref{fig:color-loss} shows that artistic control over colors is possible by providing a custom palette.

\paragraph*{Impact of multiple style images}

\begin{figure}[ht]
    \begin{tabular}{ccc}
    1 style image & 2 style images & 3 style images \\
    \includegraphics[width=0.275\linewidth]{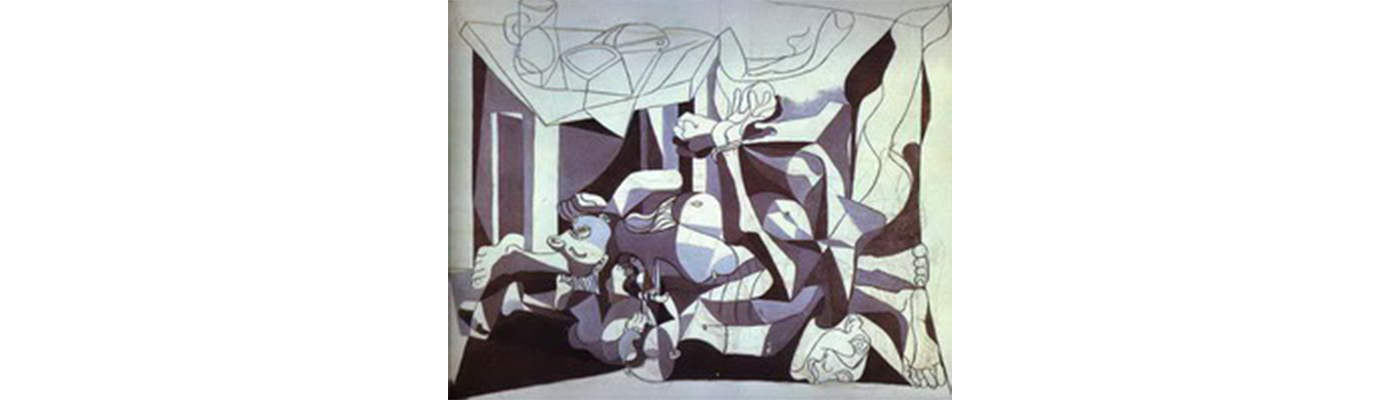} &
    \includegraphics[width=0.275\linewidth]{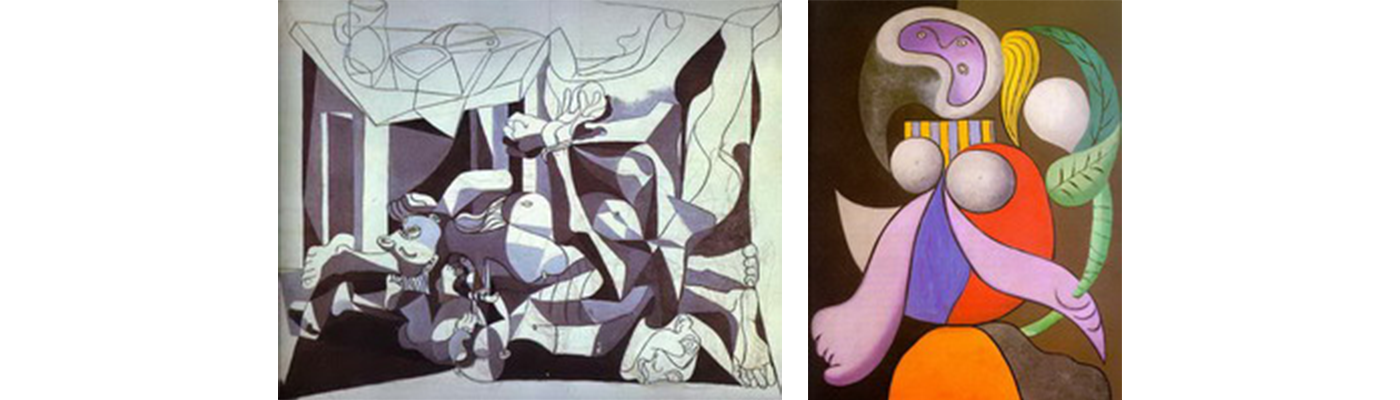} &
    \includegraphics[width=0.275\linewidth]{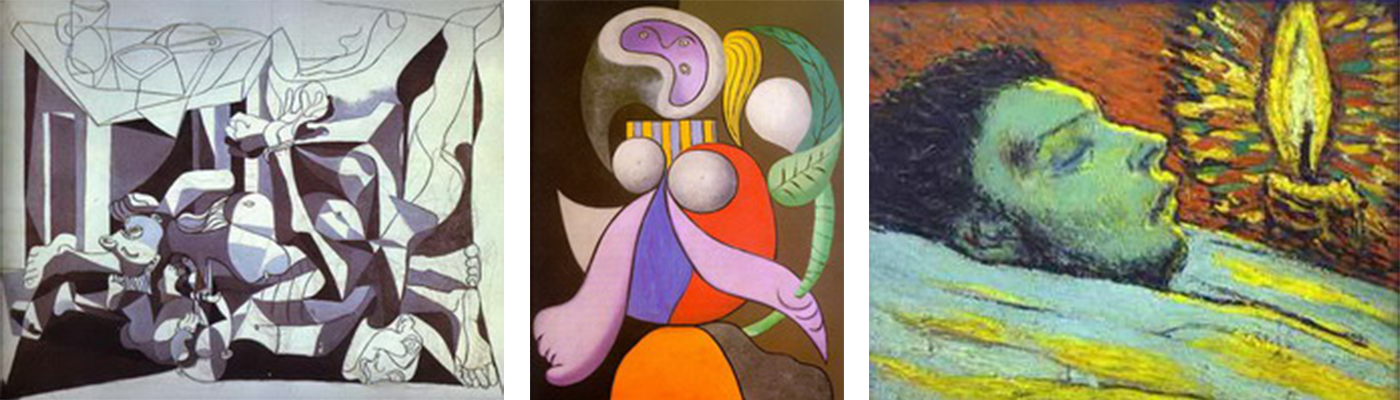} \\
    \includegraphics[width=0.275\linewidth]{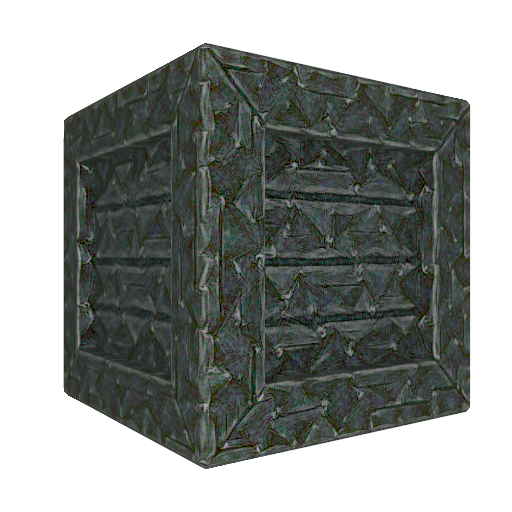} &
    \includegraphics[width=0.275\linewidth]{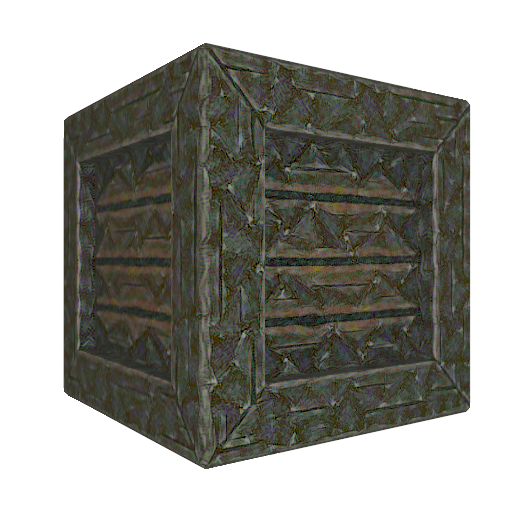} &
    \includegraphics[width=0.275\linewidth]{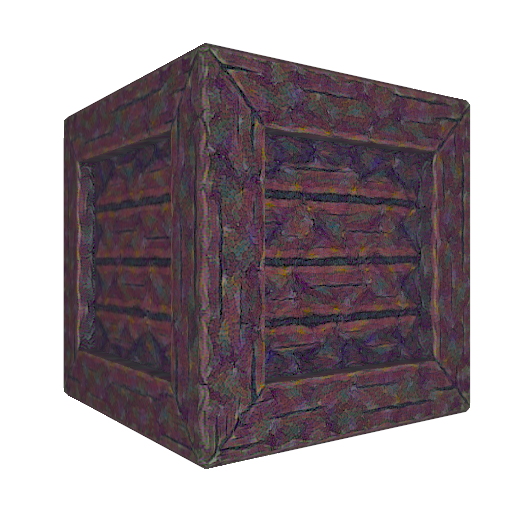} \\
    \includegraphics[clip,trim=120 120 120 120,width=0.275\linewidth]{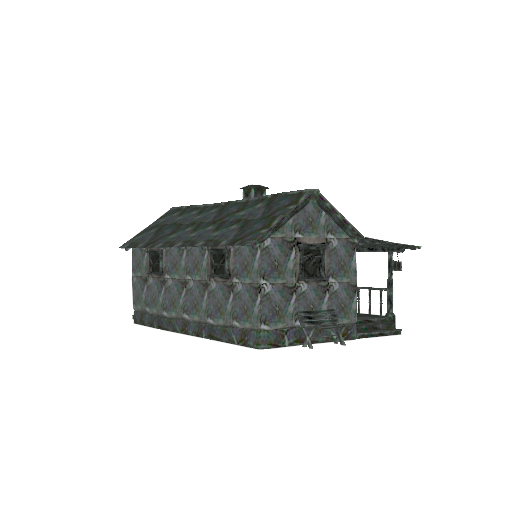} &
    \includegraphics[clip,trim=120 120 120 120,width=0.275\linewidth]{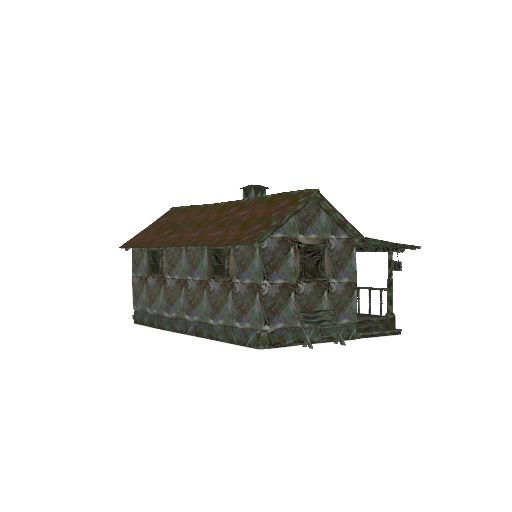} &
    \includegraphics[clip,trim=120 120 120 120,width=0.275\linewidth]{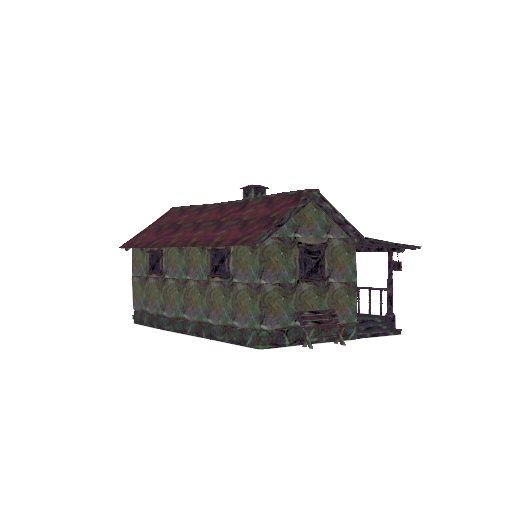} \\
    \end{tabular}
    \centering
    \caption{
        \label{fig:multiple-images}
        Additional style images affect the optimization result as there are a larger number of feature vectors to compare against in the NNFM loss.
    }
\end{figure}

The nearest-neighbor feature matching loss extends naturally to a larger number of style images simply through concatenation as shown in \autoref{sec:method}. In \autoref{fig:multiple-images}, we show the impact that additional images have on NNFM-based style transfer. As additional style images are included, the style transfer appearance changes and new patterns and colors can appear. In these results, we used the first style image as the background and for color palette extraction.

\section*{Conclusion and Future Work}

We have presented a texture stylization method for 3D assets based on CLIP-ResNet50. The nearest-neighbor feature matching loss learns style better than a Gram matrix method, and a CLIP-based loss extracts a more textural appearance than VGG16, which focuses on brush strokes and geometric patterns. We also extended NNFM to multiple style images and improved results and artistic control with a user-provided or automatically extracted color palette.

Image-based stylization with CLIP could be improved by exploring network architectures other than ResNet50 or disentangling content and style in the CLIP embedding. Neural style fields \cite{Michel:2021} could also be beneficial, especially for scene-level stylization, as texture coordinates might not be optimal (\autoref{fig:teaser}). Learning to stylize BRDF lobes, for toon shading for example, also presents an interesting opportunity for novel research.

\section*{Acknowledgements}

We thank Philipp Fischer, Guilin Liu, Jacob Munkberg, Timo Roman, Towaki Takikawa and Kangxue Yin for insightful discussions, guidance and feedback. We additionally thank the creators of the following assets; Spot and Bob from Keenan Crane, the hibiscus bush from Jacob Munkberg, the wooden box from Turbosquid user JeanSouza, the cottage model from user nukemut and the elephant drawing from publicdomainpictures.net.

\appendix

\vspace{-2mm}
\bibliographystyle{eg-alpha-doi}
\bibliography{paper}

\end{document}



\maketitle
\begin{abstract}
We present a method for transferring the style from a set of images to a 3D object. The texture appearance of an asset is optimized with a differentiable renderer in a pipeline based on losses using pretrained networks. More specifically, we utilize a nearest-neighbor feature matching loss with CLIP-ResNet50 to extract the style from images. We show that a CLIP-based loss provides a different appearance over a VGG-based loss by focusing more on texture over geometric shapes. Additionally, we extend the loss to support multiple images and enable loss-based control over the color palette combined with automatic color palette extraction from style images. 

\ccsdesc[300]{Computing methodologies~Appearance and texture representations}
\ccsdesc[300]{Computing methodologies~Rasterization}
\ccsdesc[300]{Computing methodologies~Supervised learning by regression}
\ccsdesc[100]{Applied computing~Media arts}

\end{abstract}

\section{Optimization}

In this section, we outline the details of our optimization process. All our results were produced on V100 GPUs where optimizing a single model on one GPU took three hours while using a node of eight GPUs brought the time down to 20 minutes per asset. The code was implemented in PyTorch \cite{PyTorch} and for differentiable rendering we used nvdiffrast from Laine et al. \cite{Laine:2020}

We used batch size 8 with learning rate $10^{-2}$. Unless otherwise mentioned, each batch element was randomized with camera distances based on the model and angles randomized in the sphere around the model. The power of the point light in the scene was fixed to $2$ and its position randomized in the hemisphere with radiuses in the range of $\lbrack3, 5\rbrack$. We also used a Phong material model with roughness set to 2. The render output resolution from the differentiable renderer was set to $512 \times 512$ and the texture resolution was set to $1024 \times 1024$ unless mentioned otherwise.

For our main results using CLIP-ResNet50, we use the following loss weights; $\lambda_{nnfm} = 10^4, \lambda_{content} = 22$ and the color loss weight $\lambda_{color}$ in the beginning of optimization was set to $2000$ but then reduced during the training duration. For other results, we had to tune the loss balance to adjust for missing losses or varying loss magnitudes. For VGG NNFM, we used $\lambda_{nnfm} = 200, \lambda{1.0}$ and the initial weight for the color loss was also $2000$. 

\section{Additional results}

Here, we show additional results that might give more insight into how different training aspects affect the optimization result.

\begin{figure}
\small
\begin{tabular}{ccc}
Style image & Texture Optim. & Diff. Rendering \\
\includegraphics[width=0.3\linewidth]{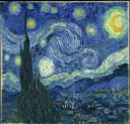} &
\includegraphics[clip,trim=50 50 50 50,width=0.3\linewidth]{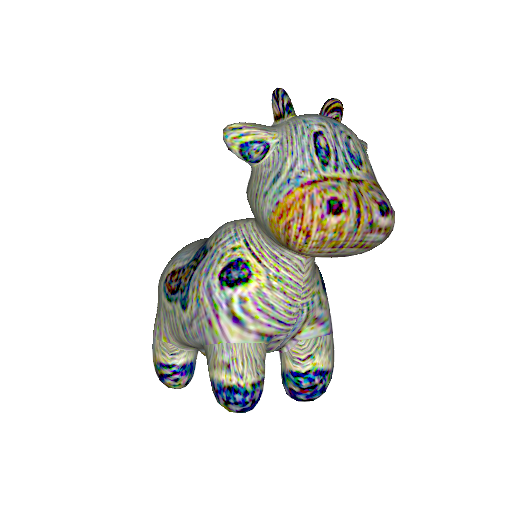} &
\includegraphics[clip,trim=50 50 50 50,width=0.3\linewidth]{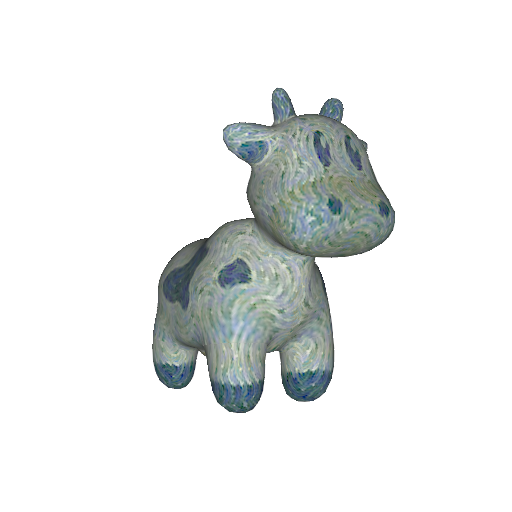} \\
\end{tabular}
\centering
\caption{
        \label{fig:direct-texture}
        Applying losses directly to the textures are another option. However, optimizing this way does not reduce discontinuities in the texture coordinates. Seams in the texture mapping are therefore more visible.
}
\end{figure}

\paragraph*{Direct texture optimization}

The differentiable renderer allows backpropagating image-space losses through the renderer to the 2D texture applied on the mesh. However, image-space losses could also be directly computed for the texture. In \autoref{fig:direct-texture}, we show a comparison of the two approaches. Differentiable rendering allows optimizing textures while keeping learned patterns consistent and scaled similarly across seams. Discontinuities are more visible when optimizing textures directly (see the front leg of the cow).

\begin{figure}
\small
\begin{tabular}{cccc}
Style image & $r=1$ & $r=2$ & $r=4$ \\
\includegraphics[width=0.2\linewidth]{images/transfer-comparison/river/river.jpg} &
\includegraphics[clip,trim=100 100 100 100,width=0.2\linewidth]{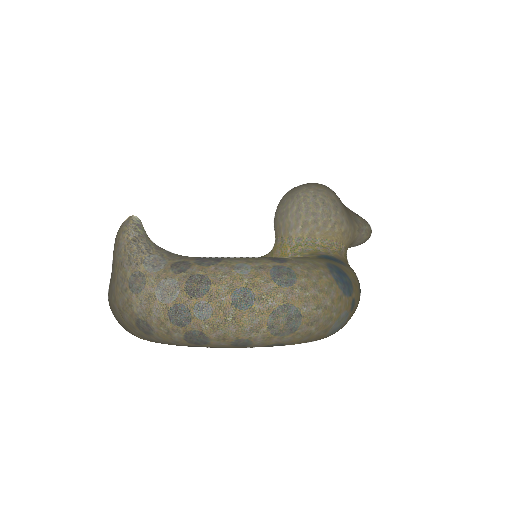} &
\includegraphics[clip,trim=100 100 100 100,width=0.2\linewidth]{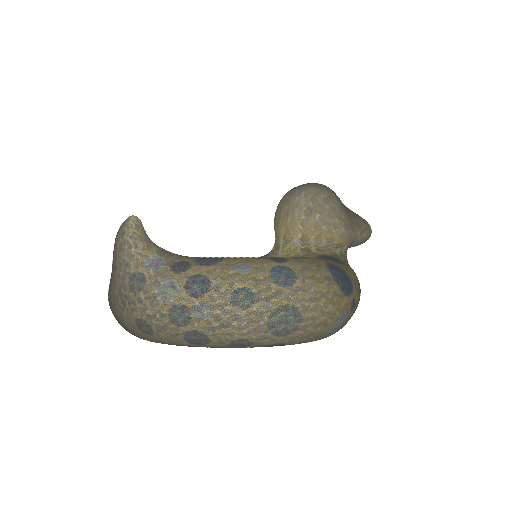} &
\includegraphics[clip,trim=100 100 100 100,width=0.2\linewidth]{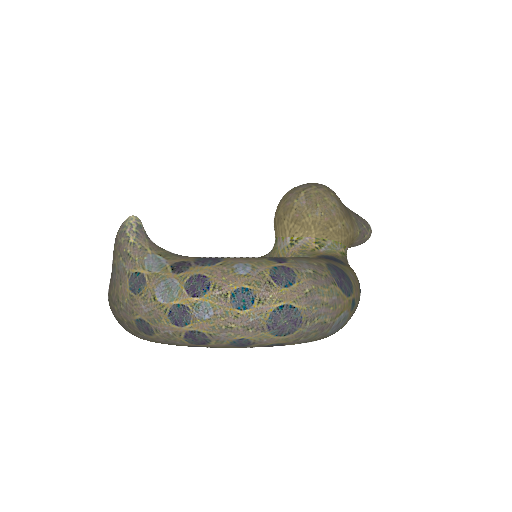} \\
\includegraphics[width=0.2\linewidth]{images/pipeline/style_image.png} &
\includegraphics[clip,trim=50 50 50 50,width=0.2\linewidth]{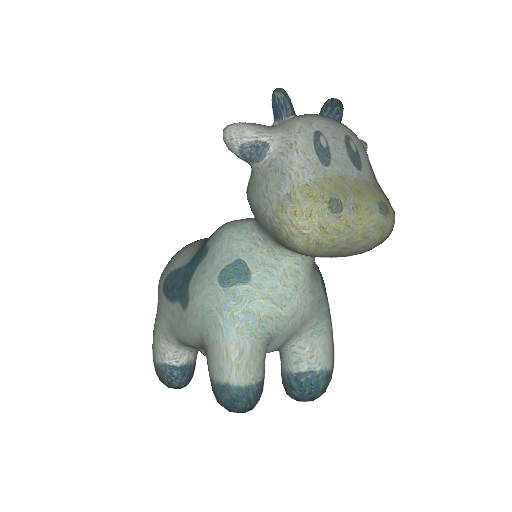} &
\includegraphics[clip,trim=50 50 50 50,width=0.2\linewidth]{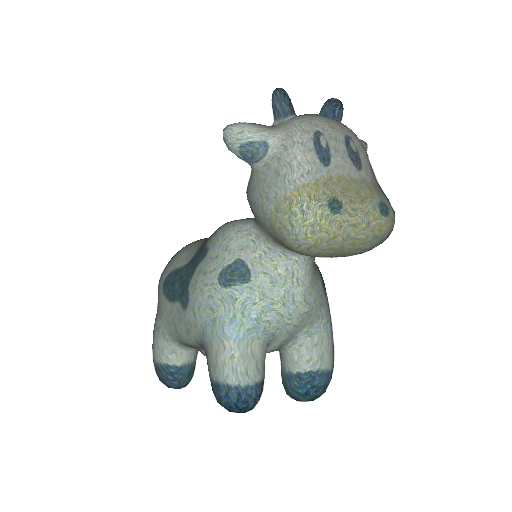} &
\includegraphics[clip,trim=50 50 50 50,width=0.2\linewidth]{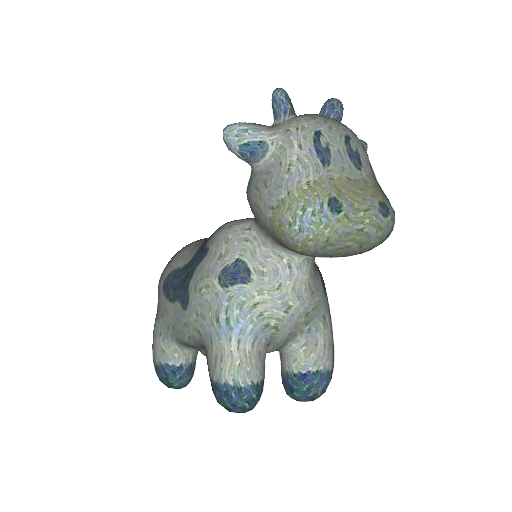} \\
\end{tabular}
\centering
\caption{
        \label{fig:camera-distance}
        Camera distance plays an important part in the size of the patterns found on the object. Here we compare different fixed camera distances $r$ used during optimization while still rotating in the sphere around the object.
}
\end{figure}

\paragraph*{Camera distance variation}

Camera distance plays a significant role in the size of patterns optimized into the texture. In \autoref{fig:camera-distance}, we show how different camera distances impact the resulting texture. Intuitively, closer camera distances result in smaller patterns while increased distance results in larger patterns. Finding the correct camera distance for a specific style requires some experimentation. 

\begin{figure}
\small
\begin{tabular}{cccc}
$64 \times 64$ & $128 \times 128$ & $256 \times 256$ & $512 \times 512$ \\
\includegraphics[clip,trim=50 50 50 50,width=0.2\linewidth]{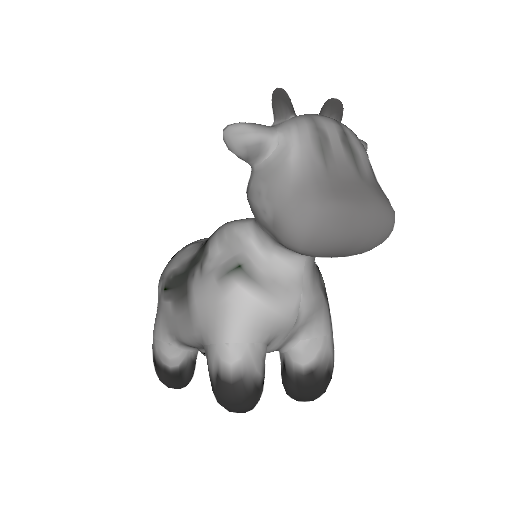} &
\includegraphics[clip,trim=50 50 50 50,width=0.2\linewidth]{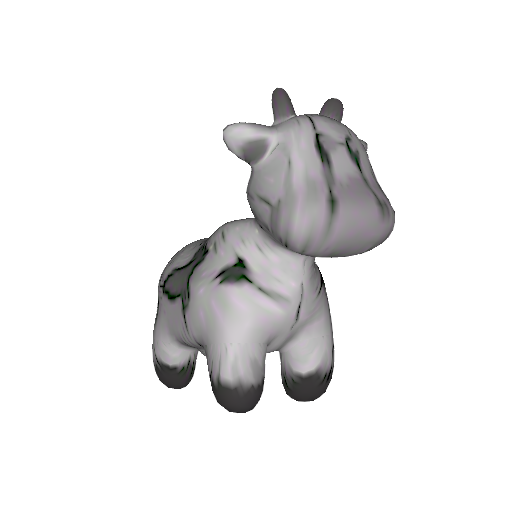} &
\includegraphics[clip,trim=50 50 50 50,width=0.2\linewidth]{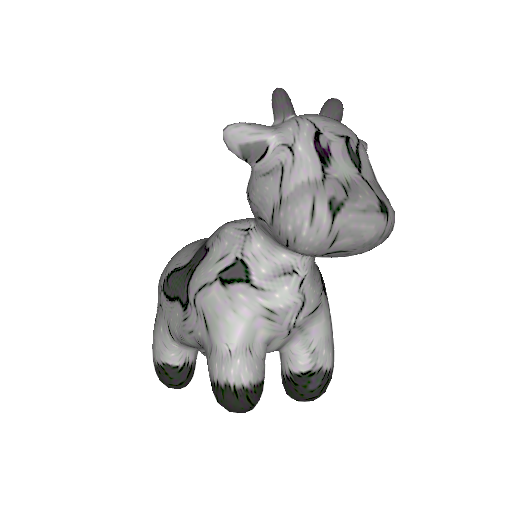} &
\includegraphics[clip,trim=50 50 50 50,width=0.2\linewidth]{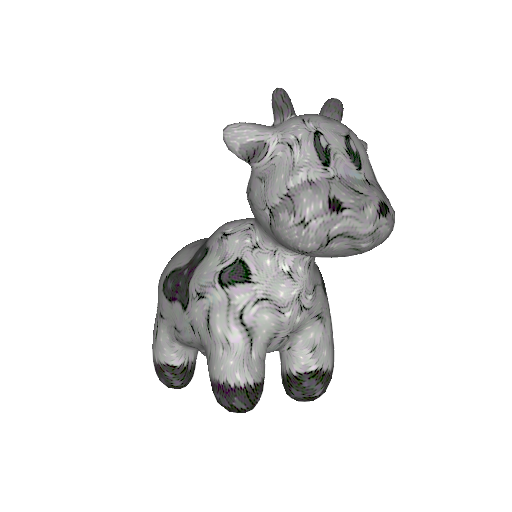} \\
\includegraphics[clip,trim=50 50 50 50,width=0.2\linewidth]{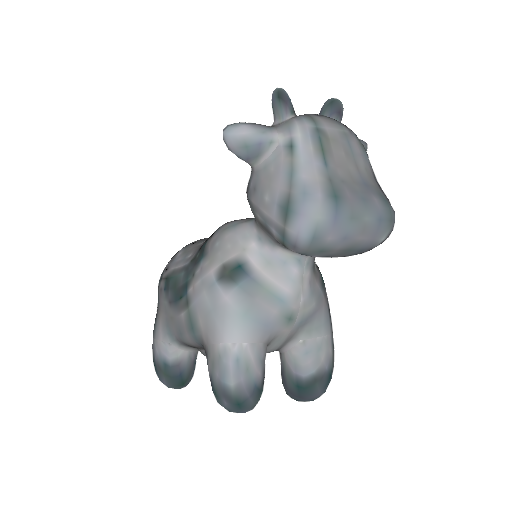} &
\includegraphics[clip,trim=50 50 50 50,width=0.2\linewidth]{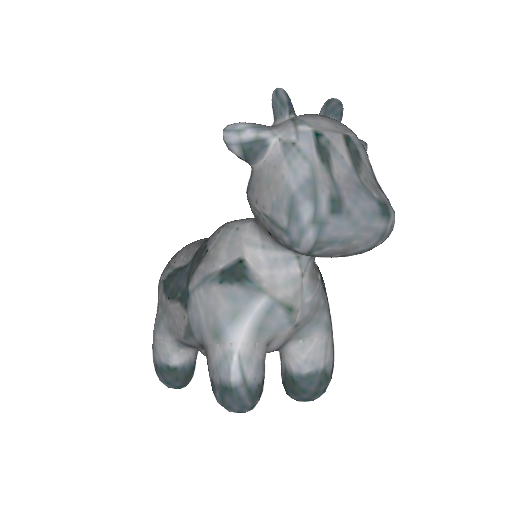} &
\includegraphics[clip,trim=50 50 50 50,width=0.2\linewidth]{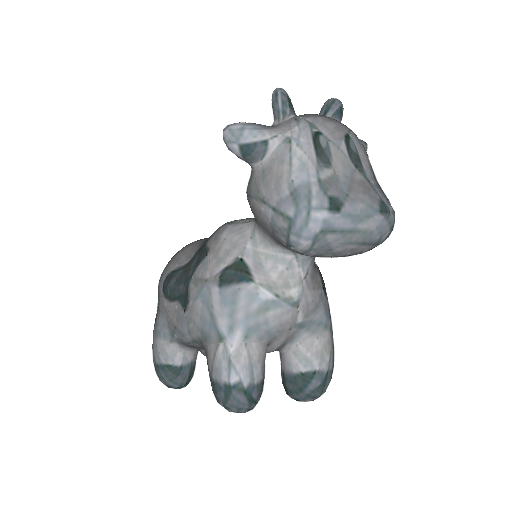} &
\includegraphics[clip,trim=50 50 50 50,width=0.2\linewidth]{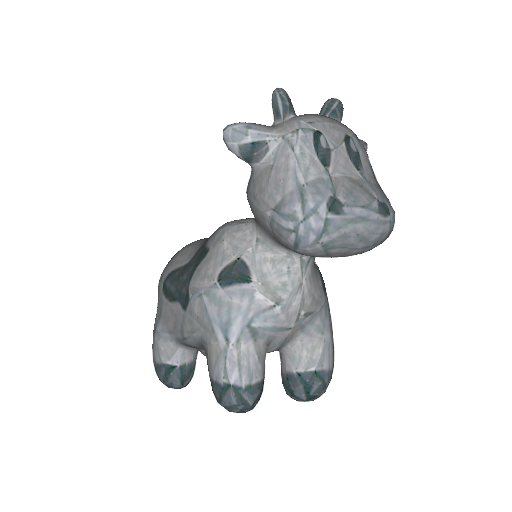} \\
\includegraphics[clip,trim=50 50 50 50,width=0.2\linewidth]{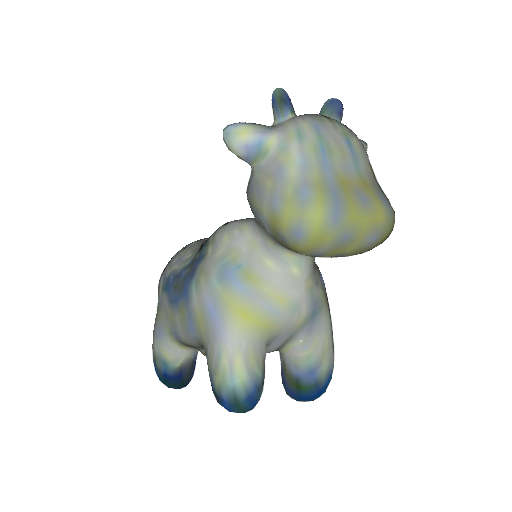} &
\includegraphics[clip,trim=50 50 50 50,width=0.2\linewidth]{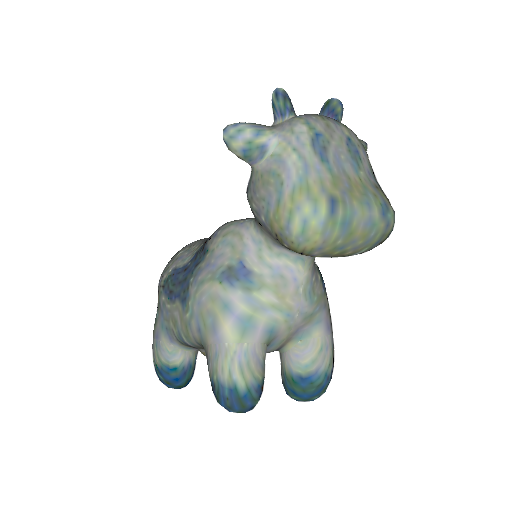} &
\includegraphics[clip,trim=50 50 50 50,width=0.2\linewidth]{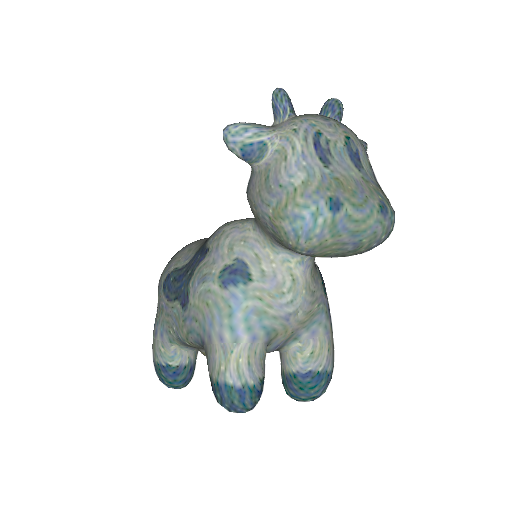} &
\includegraphics[clip,trim=50 50 50 50,width=0.2\linewidth]{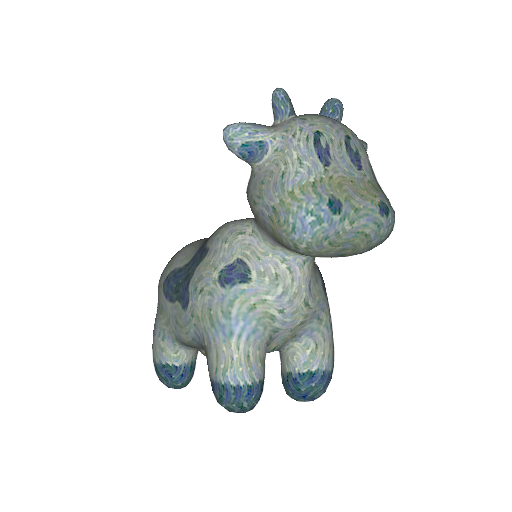} \\
\end{tabular}
\centering
\caption{
        \label{fig:texture-resolution}
        Results using various texture resolutions.
}
\end{figure}

\paragraph*{Texture resolution}

In \autoref{fig:texture-resolution}, we show the impact of texture resolution on the stylization. 

Additionally, we tried optimizing a hierarchy of textures at different resolutions. Optimization would begin at the lowest resolution and the resolution would be gradually increased. Whenever the resolution was increased, we upsampled the current lower resolution texture to the new resolution and continued optimization. However, this provided minimal benefit in most cases and finding adequate hyperparameters became too challenging. But we believe there might be something interesting to discover if further investigations are made. 

\paragraph*{Teaser image}

\begin{figure*}[ht]
 \includegraphics[width=1.0\linewidth]{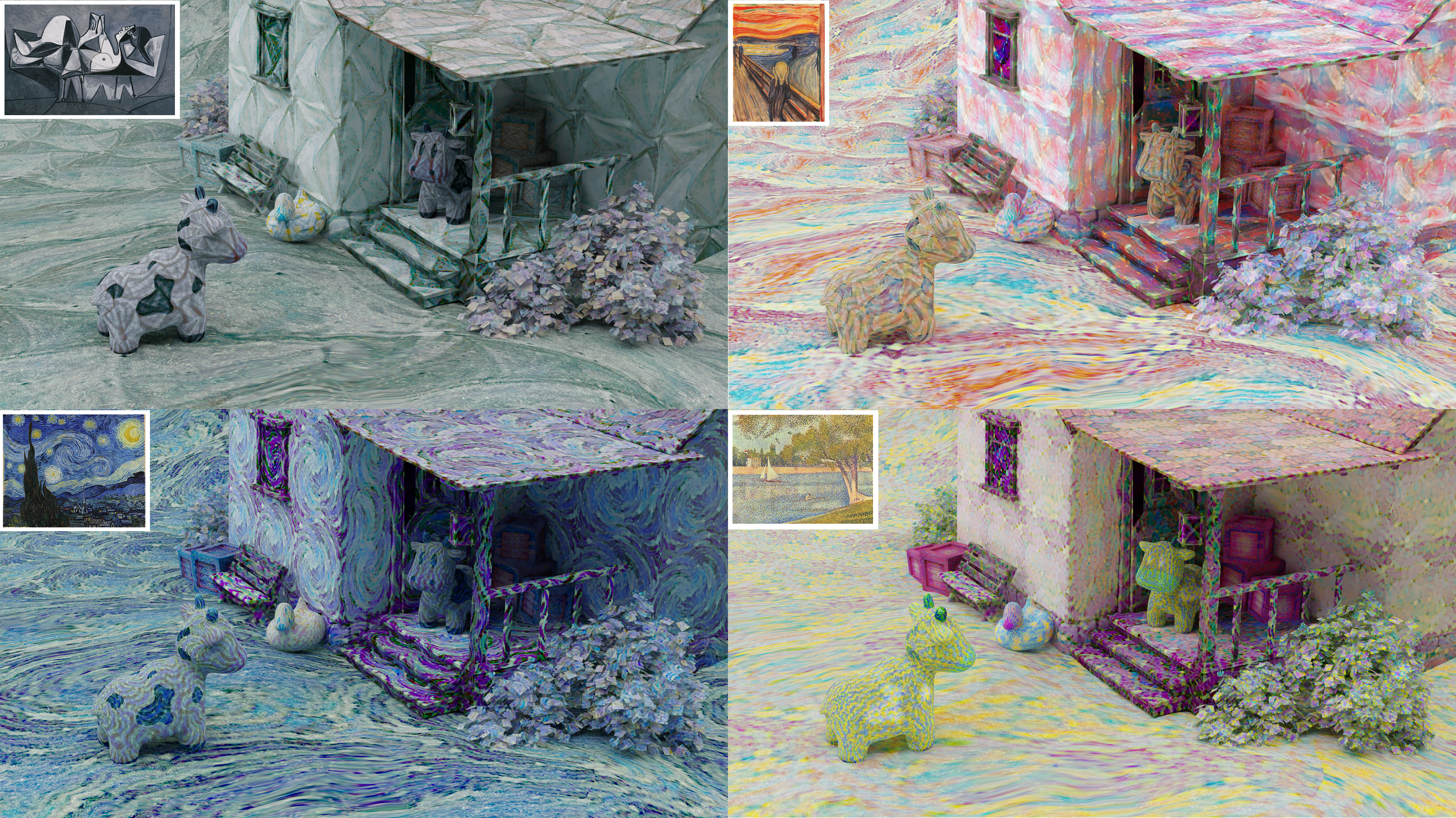}
 \centering
  \caption{Uncropped renders of the teaser figure scene. See our website for a video.}
\label{fig:teaser-web}
\end{figure*}

The teaser image in the main paper, also shown in \autoref{fig:teaser-web}, was created by stylizing objects individually using our CLIP-based NNFM method combined with content and color palette losses. The textures were then saved and added to the correct objects in Blender 3.0 where we layed out the scene. To account for our addition of the style texture and the original texture, we built a shader graph for each object performing this addition and adjusted some of the texture appearances through some manual tweaking of the blending. The scene was then rendered using Cycles. 

The teaser unfortunately also reveals a challenge with independent optimization of objects. Varying object sizes and texture coordinates makes it difficult to get similarly sized patterns appearing on scene objects. One could instead learn to map position to texture instead and optimize entire scenes simultaneously.

\begin{figure*}[ht]
\small
\begin{tabular}{ccccccc}
Style image & Original & CLIP Embedding & VGG Gram & CLIP Gram & VGG NNFM & CLIP NNFM \\
\includegraphics[clip,trim=0 0 0 0,width=0.12\linewidth]{images/loss-ablation/elephant-line-art-drawing.png} &
\includegraphics[clip,trim=50 50 50 50,width=0.12\linewidth]{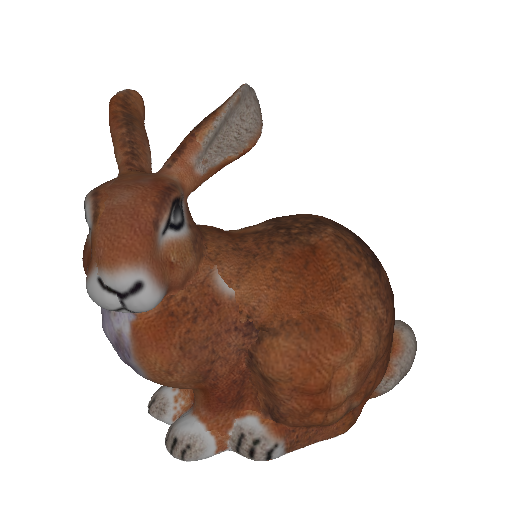} &
\includegraphics[clip,trim=50 50 50 50,width=0.12\linewidth]{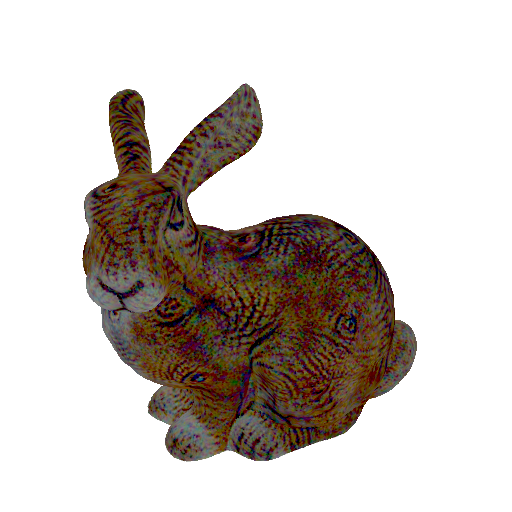} &
\includegraphics[clip,trim=50 50 50 50,width=0.12\linewidth]{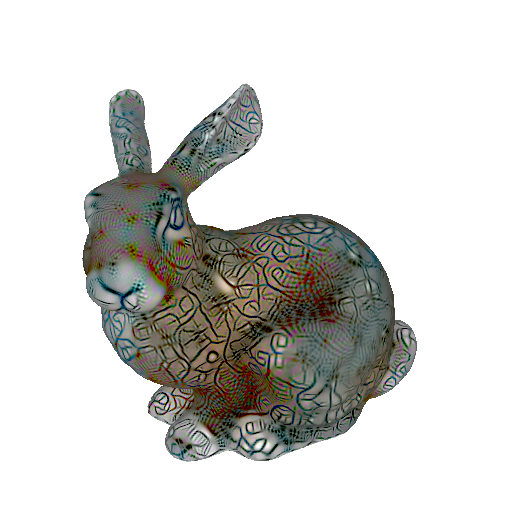} &
\includegraphics[clip,trim=50 50 50 50,width=0.12\linewidth]{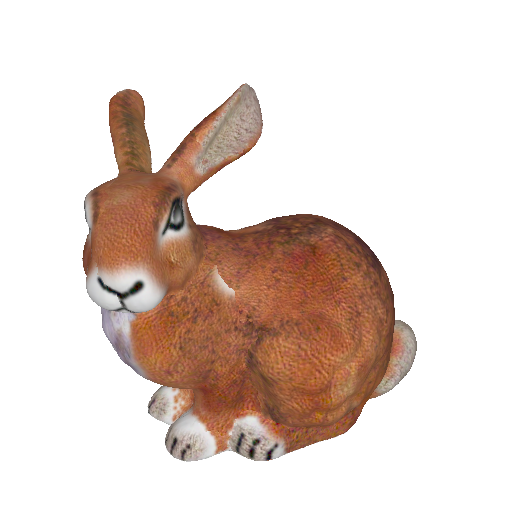} &
\includegraphics[clip,trim=50 50 50 50,width=0.12\linewidth]{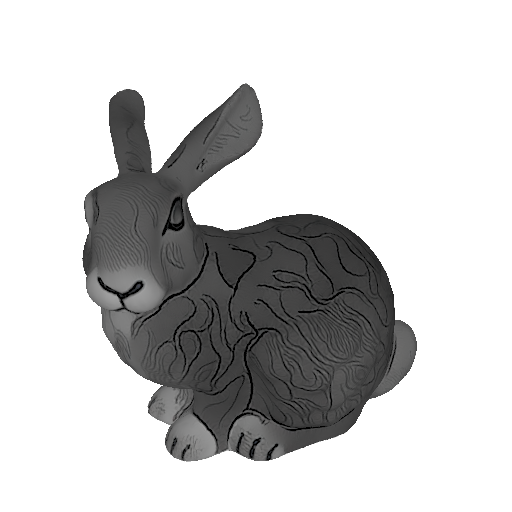} &
\includegraphics[clip,trim=50 50 50 50,width=0.12\linewidth]{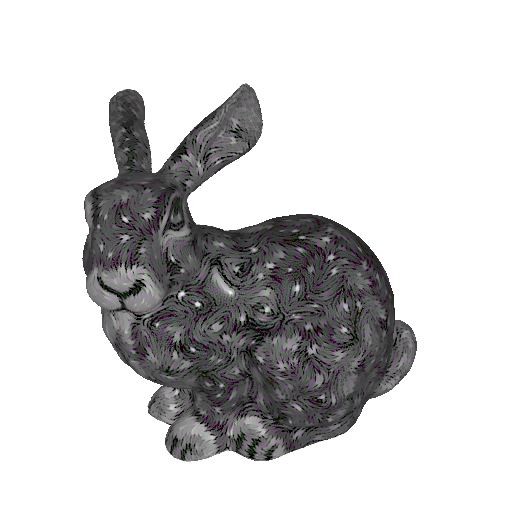} \\
\includegraphics[clip,trim=0 0 0 0,width=0.12\linewidth]{images/pipeline/style_image.png} &
\includegraphics[clip,trim=50 50 50 50,width=0.12\linewidth]{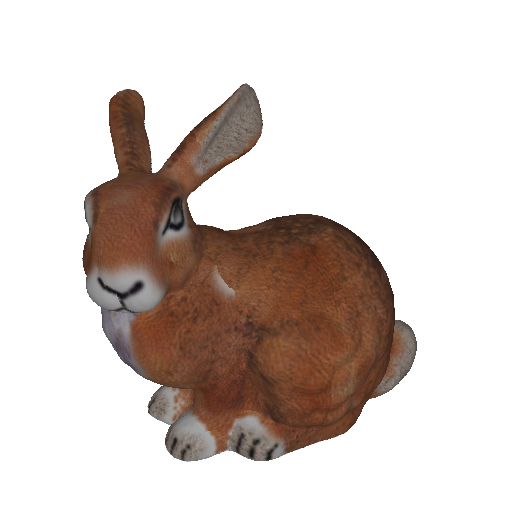} &
\includegraphics[clip,trim=50 50 50 50,width=0.12\linewidth]{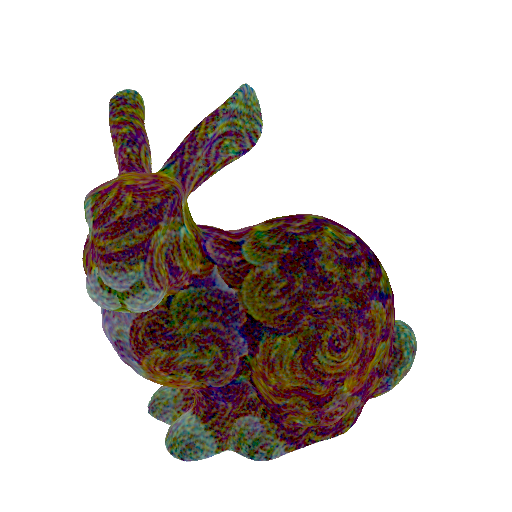} &
\includegraphics[clip,trim=50 50 50 50,width=0.12\linewidth]{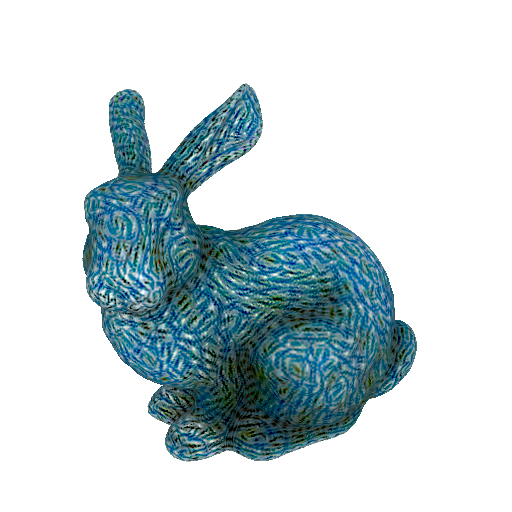} &
\includegraphics[clip,trim=50 50 50 50,width=0.12\linewidth]{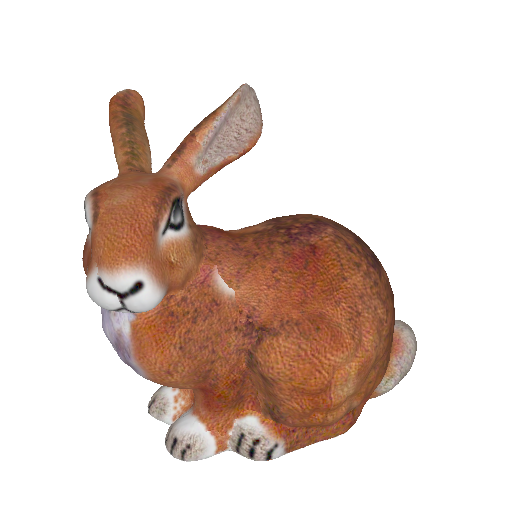} &
\includegraphics[clip,trim=50 50 50 50,width=0.12\linewidth]{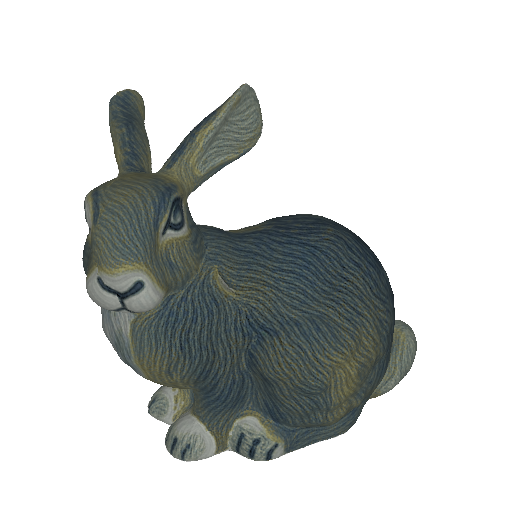} &
\includegraphics[clip,trim=50 50 50 50,width=0.12\linewidth]{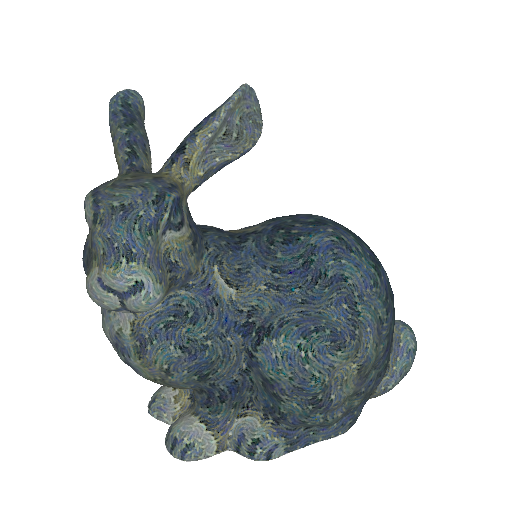} \\
\includegraphics[clip,trim=0 0 0 0,width=0.12\linewidth]{images/color-loss/picasso/style.png} &
\includegraphics[clip,trim=50 50 50 50,width=0.12\linewidth]{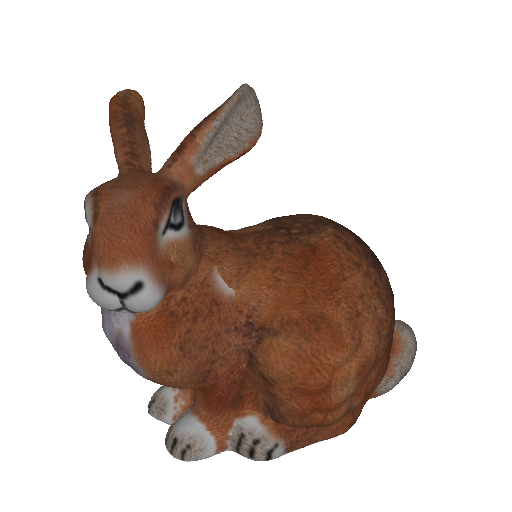} &
\includegraphics[clip,trim=50 50 50 50,width=0.12\linewidth]{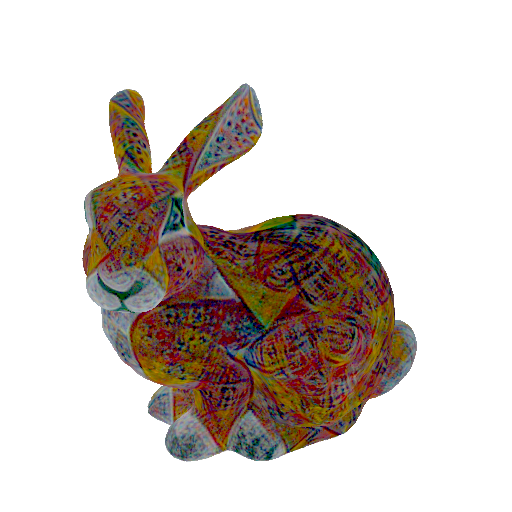} &
\includegraphics[clip,trim=50 50 50 50,width=0.12\linewidth]{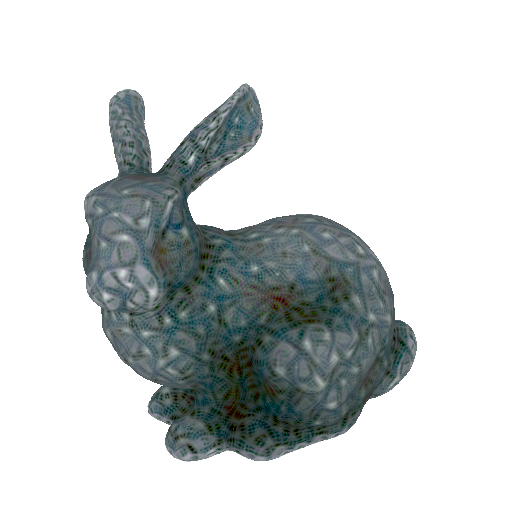} &
\includegraphics[clip,trim=50 50 50 50,width=0.12\linewidth]{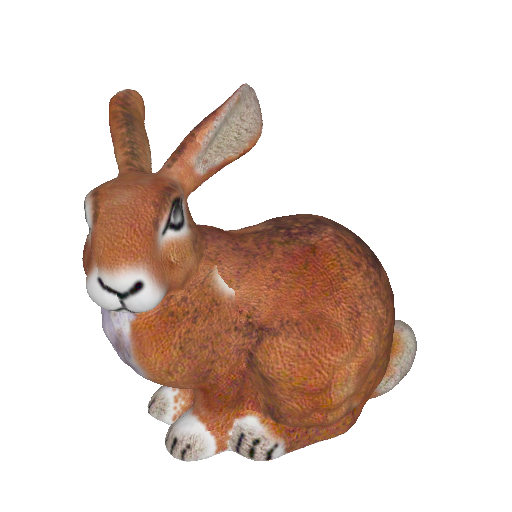} &
\includegraphics[clip,trim=50 50 50 50,width=0.12\linewidth]{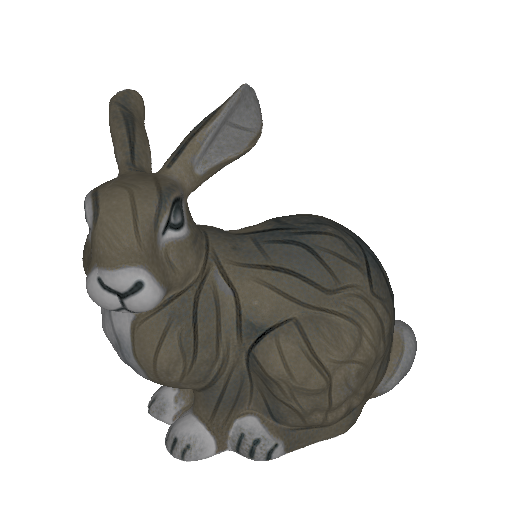} &
\includegraphics[clip,trim=50 50 50 50,width=0.12\linewidth]{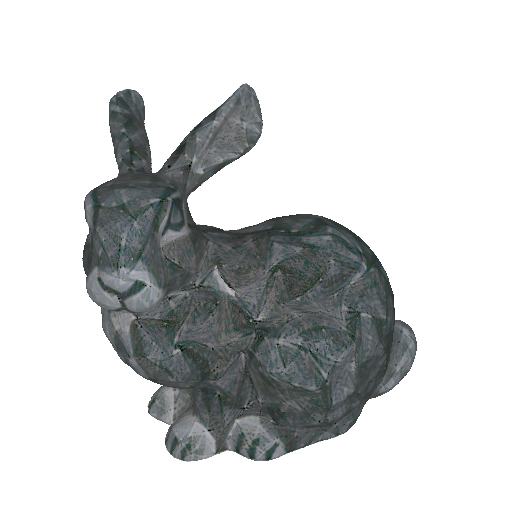} \\
\end{tabular}
\centering
\caption{
        \label{fig:transfer-comparison-supp}
        Additional comparisons of different approaches on a Stanford bunny from the Stanford University Computer Graphics Laboratory. As seen here, the CLIP Gram approach did not work most of the time. 
}
\end{figure*}

\appendix

\bibliographystyle{eg-alpha-doi}
\bibliography{supplemental}